\documentclass[12pt]{article}
\usepackage{amsmath,amssymb,amsfonts}
\usepackage{graphicx}
\usepackage{hyperref}
\hypersetup{colorlinks,linkcolor={blue},citecolor={blue},urlcolor={red}} 
\usepackage{cite}
\usepackage[english]{babel}
\usepackage{bm}
\usepackage[utf8]{inputenc}
\usepackage{listings}
\usepackage[T1]{fontenc}
\usepackage{lmodern}
\usepackage{caption}
\usepackage[normalem]{ulem}   
\usepackage{floatrow}
\usepackage{subcaption}
\usepackage{mathrsfs}
\usepackage{enumerate}
\usepackage[normalem]{ulem}
\usepackage{color}
\usepackage{array}
\usepackage[dvipsnames]{xcolor}     

\begin{document}

\title{\hfill\mbox{\small}\\[-1mm]
\hfill~\\[0mm]
       \textcolor{orange}{\textbf{Multi-component Dark Matter from an $SU(2)_{D}\to Z_{3}^{D}\times  Z_{2}^{{\rm acc}}$ scenario }}        }
\date{}
\author{
\\[1mm] 
Carlos Alvarado$^{1\,}$\footnote{E-mail: {\tt calvarado@vassar.edu}}~,
Alfredo Aranda$^{2,3}$\footnote{E-mail: {\tt fefo@ucol.mx }}~, Cesar Bonilla$^{4\,}$\footnote{E-mail: {\tt cesar.bonilla@ucn.cl }}~,\\
and Carlos Ramos Portalatino$^{5\,}$\footnote{E-mail: {\tt ramosport.carlos@ufl.edu }}
\\[1mm]
	\textit{\small $^1 $Physics and Astronomy Department, Vassar College,}\\
	\textit{\small Box 745, Poughkeepsie, NY 12604, U.S.A.}\\[3mm]	
	\textit{\small $^2$Dual CP Institute of High Energy Physics,}\\
	\textit{\small C.P.~28045, Colima, M\'exico}\\[3mm]
	\textit{\small $^3 $Facultad de Ciencias, Universidad de Colima,}\\
	\textit{\small C.P.~28045, Colima, M\'exico}\\[3mm]
	\textit{\small $^4 $Departamento de F\'isica, Universidad Cat\'olica del Norte,}\\
	\textit{\small Avenida Angamos 0610, Casillas 1280, Antofagasta, Chile}\\[3mm]
        \textit{\small $^5 $Facultad de Ciencias, Universidad Nacional de Ingenier\'ia,}\\ \textit{\small Avenida T\'upac Amaru 210, R\'imac 15333, Lima, Per\'u}\\[3mm]
    }
\maketitle
\vspace{0.5cm}

\vspace*{-1.0cm}
\begin{abstract}
\noindent

A multi-component weakly-interacting dark matter scenario is analyzed, where the candidate dark matter states arise from the $SU(2)_{D}\to Z_{3}^{D}\times Z_{2}^{\text{acc}}$ symmetry breaking triggered by a VEV-developing scalar in the $J=3/2$ representation. While the discrete symmetries guarantee at least two stable states, a closer look at the mass orderings of the contrived dark sector reveals the existence of a third dark matter candidate, whose stability effectively results from the tight phase space in the dark boson sector. Regions of masses and couplings where the three dark states furnish a comparable contribution to the observed relic abundance are determined numerically.
\end{abstract}
\thispagestyle{empty}
\vfill
\setcounter{page}{1}

\section{Introduction\label{sec:intro}}

The identity of the long-ago observed dark matter in the skies persists as a puzzling mystery for the cosmology, astrophysics, and particle physics communities alike. Initially detected by gravitational means and later by its imprint on the Cosmic Microwave Background spectrum, dark matter (DM) has been shown to typically permeate galactic halos and constitute about 26\% of the total energy budget of the universe.

Attempts to provide a successful answer to this mystery have led these communities to entertain the possibility that dark matter is a yet undiscovered fundamental particle (or set of particles) with electroweak interactions. This entire line of thought, nicknamed the Weakly Interactive Massive Particle (WIMP) paradigm, has seen plenty of literature produced. The simplest cases have been considered in the past, ranging from dark matter candidates entirely neutral under the Standard Model (SM) gauge interactions (singlet DM \cite{Silveira:1985rk, Burgess:2000yq, McDonald:1993ex, Cline:2013gha}) to candidates that form composite states (bound-state DM \cite{Alves:2009nf, Asadi:2016ybp, vonHarling:2014kha, Aranda:2015jis}). Another direction has been brought by scenarios where DM transforms under nontrivial multiplet representations of the SM electroweak gauge group \cite{LopezHonorez:2006gr, Avila:2019hhv, Zeng:2019tlw, Ostdiek:2015aga}. This premise has been combined with the hypothesis that DM carries quantum numbers under discrete symmetries which apart from stabilizing it by preventing its decay, serve other purposes like the generation of tiny neutrino masses \cite{Belanger:2012zr, Bonilla:2016diq, Dasgupta:2019rmf, Belanger:2020hyh}.

The work presented here belongs to the latter class, where DM is made of multiple species, which carry charges under symmetry groups that extend the SM one. The DM candidates arise from scalars in a higher representation of a dark-sector $SU(2)_D$ gauge symmetry, known as the quadruplet representation ($J=3/2$ in the language of group theory), whose chosen vacuum expectation value (VEV) direction induces a remnant cyclic $Z_{3}^D$. This model was initially proposed by Borah, Ma, and Nanda~\cite{Borah:2022dbw}, whose work took simplifying limits to ease the discussion of its phenomenology. Furthermore, the authors pointed out the presence of an additional symmetry, an accidental $ Z_{2}^{{\rm acc}}$, that enriches the setup.

In this paper we revisit that model and offer a classification of the possible DM scenarios according to the number of dark candidates that contribute significantly to the observed dark matter abundance. There is a strong dependence of these scenarios on the mass relations/kinematics between all states charged under $Z_{3}^{D}\times Z_{2}^{{\rm acc}}$, and a nontrivial interplay between the two Higgs portals that offers the conditions necessary for three-component DM where each species contributes a comparable amount to the entire abundance. In addition, the direct detection prospects for the different mass orderings are discussed, and how these are modified by the multicomponent nature of DM. Likewise, we mention the impact of semi-annihilation at the selected benchmark parameter choices. For other multi-species dark matter analyses, please refer to \cite{Belanger:2020hyh, Belanger:2022esk}.

The description of the original model and notation is provided in Sec.~\ref{sec:DSB}, followed by a discussion of the mass spectrum and the DM species mass orderings in Sec.~\ref{sec:spectrum}. Once equipped with the analytical mass dependence, the classification of the scalar DM scenarios and corresponding kinematic conditions is presented in detail in Sec.~\ref{sec:scenarios}. This is followed by the numerical analysis of said scenarios with the help of automatized parameter space scans in Sec.~\ref{sec:numerical}. Afterwards, we offer our closing remarks.

\section{Dark symmetry breaking\label{sec:DSB}}

The SM gauge group is extended by a local $SU(2)_{D}$ symmetry, under which only a new $J=3/2$ multiplet $\Phi=(\phi_{3},\phi_{2},\phi_{1},\phi_{0})^{T}$ (quadruplet) transforms nontrivially. The full scalar potential $V(H,\Phi)$ includes the potential of the Higgs doublet $H=(H^{+},H^{0})$, with $H^{0}=(v_{h}+\rho_{h}+i\sigma_{h})/\sqrt{2}$,
\begin{equation}
V(H)=-\mu_{H}^{2}H^{\dag}H+\dfrac{\lambda_{H}}{2}(H^{\dag}H)^{2}~,
\end{equation}
a \textit{Higgs portal} operator parametrized by the quartic coupling $\lambda_{H\Phi}$,
\begin{equation}
V(H,\Phi)=V(\Phi)+V(H)+\lambda_{H\Phi}H^{\dag}H\sum_{k}\phi_{k}^{\dag}\phi_{k}~, \label{eq:potentialHPhi}
\end{equation}
and a rich quadruplet-only potential $V(\Phi)$, quoted in Appendix \ref{sec:appA}. The breaking of $SU(2)_{D}$ is obtained through a nonzero vacuum expectation value (VEV) of the quadruplet $\Phi$, with complex entries $\phi_{i}=v_{i}+(\rho_{i}+i\sigma_{i})/\sqrt{2}$, which has been explored in Ref.~\cite{Adulpravitchai:2009kd}. When $\Phi$ acquires a VEV in the direction
\begin{equation}
\langle \Phi\rangle=
\left(\begin{array}{c}
v_{3}/\sqrt{2} \\
0 \\
0 \\
v_{0}/\sqrt{2} \\
\end{array}\right)~,\label{eq:pattern}
\end{equation}
it gives rise to a residual $Z_{3}^{D}$ symmetry. Furthermore, Ref.~\cite{Borah:2022dbw} pointed out the existence of an accidental $Z_{2}^{{\rm acc}}$ symmetry upon $SU(2)_{D}$ breaking. Together, the full $Z_{3}^{D}\times Z_{2}^{{\rm acc}}$ symmetry implies a richer setup able to embed a multi-component DM sector. The physical fields after $SU(2)_{2}$ spontaneous symmetry breaking, but before electroweak symmetry breaking (EWSB), are
\begin{align}
\widetilde{\phi}_{R} &= s_{\theta}\rho_{0}-c_{\theta}\rho_{3}, \\
   \widetilde{\zeta} &= v_{D}+c_{\theta}\rho_{0}+s_{\theta}\rho_{3}, \\
            \phi_{I} &= s_{\theta}\sigma_{0}+c_{\theta}\sigma_{3}, \\
                \eta &= s_{\theta}\phi_{1}+c_{\theta}\phi_{2}^{*},
\label{eq:physicalPreSSB}
\end{align}
where $\theta$ is the quadruplet mixing angle. Upon EWSB, $\widetilde{\phi}_{R}$ and $\widetilde{\zeta}$ are subject to further mixing with $\rho_{h}$, resulting into threee physical CP-even scalars $h$, $\phi_{R}$, $\zeta$. The full, post-EWSB mixing matrices are listed in detail in Appendix \ref{sec:appB}. The new fields, which only extend the SM gauge and scalar sectors, are portrayed together with the SM Higgs in Table \ref{tab:fields}. 

\setlength\tabcolsep{12pt} 
\begin{table}[h!]
\centering
\begin{tabular}{|c|c|c|} 
\hline
\textbf{Field} & $\boldsymbol{Z_{3}^{D}}$ & $\boldsymbol{ Z_{2}^{\textbf{acc}}}$ \\
\hline \hline
$\eta$     & $\omega$ & 1  \\
$\phi_{I}$ & $1$      & -1 \\
$\phi_{R}$ & $1$      & -1 \\
$\zeta$    & $1$      & 1  \\
$h$        & $1$      & 1  \\
\hline
$X'$       & $\omega$ & -1 \\
$X_{3}$    & $1$      & 1 \\
\hline
\end{tabular}
\caption{Bosonic field content and quantum numbers after $SU(2)_{D}\to Z_{3}^{D}\times  Z_{2}^{{\rm acc}}$ spontaneous symmetry breaking. Here $\omega$ is the unit cubic root, $\omega=e^{2\pi i/3}$. The first five rows correspond to spin zero fields with $h$ representing the SM Higgs field. The last two rows contain the new gauge fields. See section~\ref{sec:spectrum} for details.} \label{tab:fields}
\end{table}

\section{New states mass spectrum \label{sec:spectrum}}

The VEV direction in Eq.~(\ref{eq:pattern}) results in three massive gauge bosons: a degenerate pair $X_{1,2}$ of mass $M_{1}=M_{2}$, and a slightly heavier $X_{3}$ of mass $M_{3}$,
\begin{equation}
M_{3}^{2}=\dfrac{9}{2}g_{D}^{2}v_{D}^{2},~~~~~M_{1}^{2}=\dfrac{3}{2}g_{D}^{2}v_{D}^{2}. \label{eq:xmasses}
\end{equation}
For convenience $X_{1}$ and $X_{2}$ will be combined into a complex vector pair $X'$, $X'^{*}$ with $X'=(X_{1}- iX_{2})/\sqrt{2}$ and $X'^{*}=(X_{1}+ iX_{2})/\sqrt{2}$. The dark gauge bosons do not mix with the SM vectors after EWSB because $\Phi$ does not carry electroweak numbers, nor through kinetic mixing as is the case with dark $U(1)'$s. The mass relation
\begin{equation}
M_{3}=\sqrt{3}M_{1} \label{eq:gaugeMassRel}
\end{equation}
between the new gauge bosons will be relevant when discussing mass gaps in Sec.~\ref{sec:scenarios}. The number of degrees of freedom in $\Phi$ results in the extended field content of Table~\ref{tab:fields}. There is a complex scalar $\eta$, a CP-odd scalar $\phi_{I}$, and two CP-even scalars $\phi_{R}$ and $\zeta$. The new states carry charges under $Z_{3}^{D}$ or $ Z_{2}^{{\rm acc}}$, except for $\zeta$, which mixes with the Higgs $h$ due to the portal quartic $\lambda_{H\Phi}$. For reference, the mixing matrices and Goldstone bosons will be listed in Appendix \ref{sec:appB}.

We trade a subset of the quartic couplings for the scalar physical masses,
\begin{align}
\lambda_{1} &= \dfrac{
\sqrt{(M_{h}^{2}-M_{\zeta}^{2})^{2}-\Delta}
}{4v_{D}^{2}}+\dfrac{
2M_{h}^{2}+M_{I}^{2}+2M_{\zeta}^{2}+3M_{\eta}^{2}-(M_{I}^{2}-3M_{\eta}^{2})(1-2x)
}{8v_{D}^{2}}, \\
\lambda_{2} &= \dfrac{
3\bigl( M_{\eta}^{2}-M_{I}^{2}+(M_{I}^{2}-3M_{\eta}^{2})(1-2x) \bigr)
}{8v_{D}^{2}}, \\
\lambda_{3} &= \dfrac{
3M_{\eta}^{2}-3M_{I}^{2}-(M_{I}^{2}-3M_{\eta}^{2})(1-2x)
}{4v_{D}^{2}}, \\
\lambda_{h} &= \dfrac{M_{h}^{2}+M_{\zeta}^{2}-\sqrt{(M_{h}^{2}-M_{\zeta}^{2})^{2}-\Delta}
}
{2v^{2}},
\\
  M_{R} &= \sqrt{3M_{\eta}^{2}-M_{I}^{2}}, \label{eq:scalarMassRel}
\end{align}
where $\Delta\equiv 8\lambda_{H\Phi}^{2}v_{D}^{2}v^{2}$ and $x\equiv \sin^{2}\textbf{}{(2\theta)}$ parametrizes the quadruplet mixing. 

\begin{figure}
\centering
\includegraphics[scale=1]{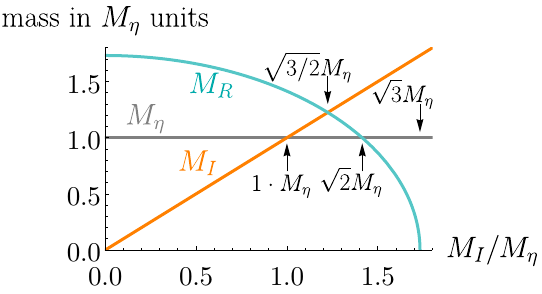}
\hspace{10mm}
\includegraphics[scale=0.7]{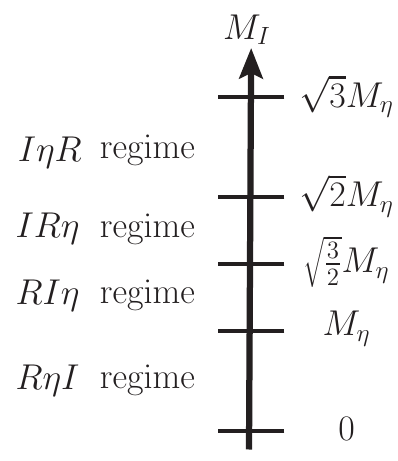}
\caption{   \textbf{Left:} compressed $\phi_{R}$, $\phi_{I}$, and $\eta$ mass spectrum from Eq.~(\ref{eq:scalarMassRel}). Masses are expressed in units of $M_{\eta}$ for illustration. \textbf{Right:} Possible mass orderings for $\phi_{R}$, $\phi_{I}$, and $\eta$ as $M_{I}$ varies at fixed $M_{\eta}$. }. \label{fig:compressed}
\end{figure}

The last of these equations is a mass relation that severely restricts the mass spectrum of the scalars with nontrivial $Z_{3}^{D}\times  Z_{2}^{{\rm acc}}$ charges. In fact, it is important to note, our relation~\eqref{eq:scalarMassRel} differs from the one obtained in Ref.~\cite{Borah:2022dbw}, where a weaker condition on those scalar masses was obtained. This is most relevant to the results presented in this paper since our stringer constraining condition implies the existence of a third scalar DM candidate, stabilized by kinematics only. At fixed $M_{\eta}$, Eq.~(\ref{eq:scalarMassRel}) not only forbids $\phi_{R}$ and $\phi_{I}$ to take masses outside the $[0,\sqrt{3}M_{\eta}]$ range, but also clusters their masses around that of $\eta$. Such compressed spectrum is portrayed in the left panel of diagram of Fig.~\ref{fig:compressed}.

As $M_{I}$ varies, with $M_{\eta}$ fixed, the $\phi_{R}$, $\phi_{I}$, and $\eta$ change their relative mass sizes, giving rise to four possible orderings. The turnover points that define these orderings are $M_{\eta}$, $\sqrt{3/2}M_{\eta}\approx 1.22M_{\eta}$, and $\sqrt{2}M_{\eta}\approx 1.41M_{\eta}$ (see right panel of Fig.~\ref{fig:compressed}). As a means to aid the classification of the various mass orderings in Sections \ref{subsec:unstab} and \ref{subsec:viable}, we adopt the following labels
\begin{align}
I\eta R~\text{regime}&~~~~~M_{I}>M_{\eta}>M_{R}~~~\text{where}~~~~~\sqrt{2}M_{\eta}<M_{I}<\sqrt{3}M_{\eta}, \notag \\
IR\eta~\text{regime}&~~~~~M_{I}>M_{R}>M_{\eta}~~~\text{where}~~~~~\sqrt{3/2}M_{\eta}<M_{I}<\sqrt{2}M_{\eta}, \notag\\
 RI\eta~\text{regime}&~~~~~M_{R}>M_{I}>M_{\eta}~~~\text{where}~~~~~M_{\eta}<M_{I}<\sqrt{3/2}M_{\eta}, \notag\\
R\eta I~\text{regime}&~~~~~M_{R}>M_{\eta}>M_{I}~~~\text{where}~~~~~0<M_{I}<M_{\eta}.
\label{eq:regimeMaster}
\end{align}
These orderings, except for the $R\eta I$ one, result in narrow ranges for the $M_{\eta}/M_{I}$ ratio.

\section{Dark sector scenarios\label{sec:scenarios}}

\begin{figure}
\centering
\includegraphics[scale=0.5]{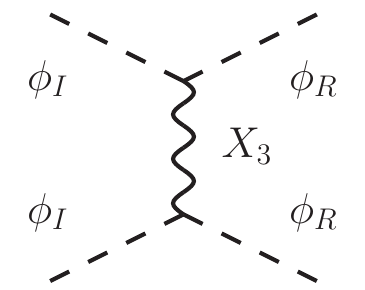}
\vspace{5mm}
\includegraphics[scale=0.5]{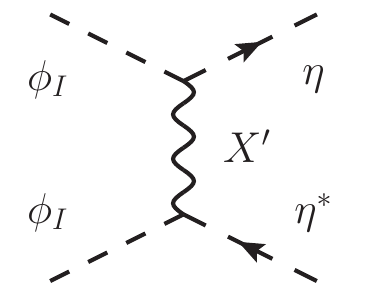}
\hspace{5mm}
\includegraphics[scale=0.5]{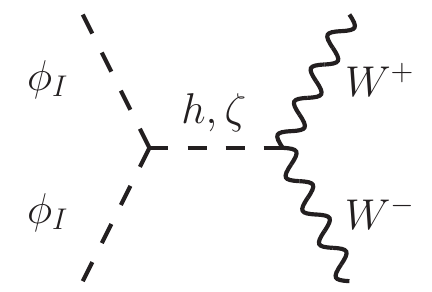}
\hspace{5mm}
\includegraphics[scale=0.5]{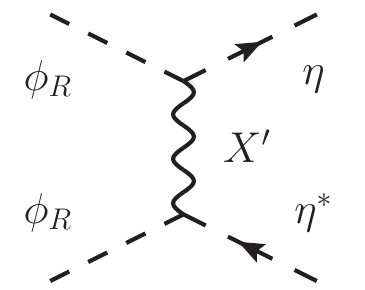}
\hspace{5mm}
\includegraphics[scale=0.5]{all_graphs/feynman_diagrams/Feyn_phiRphiR_to_etaetac_tChann_Xp.pdf}
\vspace{5mm}
\includegraphics[scale=0.5]{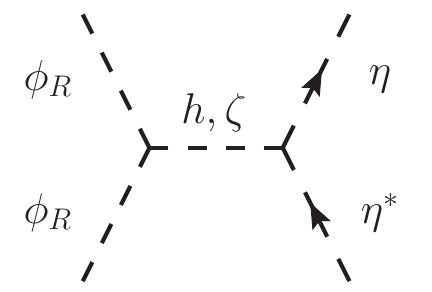}
\hspace{5mm}
\includegraphics[scale=0.5]{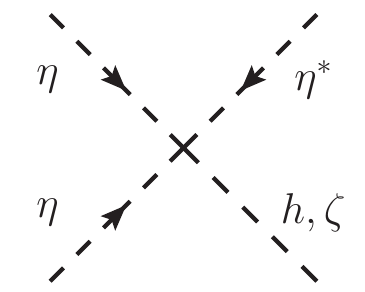}
\hspace{5mm}
\includegraphics[scale=0.5]{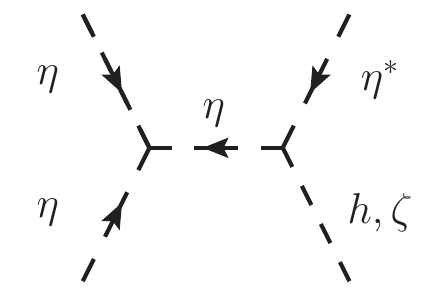}
\hspace{5mm}
\includegraphics[scale=0.5]{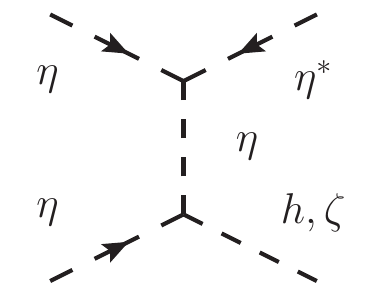}
\hspace{5mm}
\includegraphics[scale=0.5]{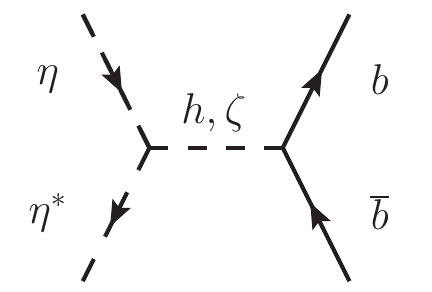}
\hspace{5mm}
\caption{Several 2-to-2 self-annihilation and semi-annihilation (for $\eta$ only) amplitudes for the dark scalars.}
\label{fig:annDiagrams}
\end{figure}

The $ Z_{2}^{{\rm acc}}$-carrying $\phi_{R}$, $\phi_{I}$ states undergo self-annihilation into other dark states (DM interconversion) or into SM species, but only through the Higgs portal. Due to this, the annihilation into the SM occurs preferentially into $W$ bosons and $b$ quarks. Meanwhile, the $Z_{3}^D$-carrying $\eta$ participates in semi-annihilation on top of its self-annihilation modes, provided the kinematics allows it. A subset of all possible 2-to-2 dark scalar annihilation topologies relevant to our benchmark choices in Section \ref{sec:numerical} is portrayed in Fig. \ref{fig:annDiagrams}.

Before reviewing the instability conditions in the dark sector, it is necessary to list the possible decay modes, which are dictated by the discrete group assignments in Table \ref{tab:fields}. Due to the small mass gaps between dark scalars, a reasonable approach to the instability conditions of all new states consists in looking at their two-body decays at tree-level. In setups where the gauge bosons are heavier than the scalar sector but still within an order of magnitude away from it, the (few) phase-space reduced three-body decays may as well be forbidden.

\subsection{Decays\label{subsec:decays}}

\begin{itemize}

\item{\textit{Decays of $\eta$.}} The complex scalar $\eta$ carries a $Z_{3}^{D}$ charge. For this reason it must include $X'$ as a daughter if unstable. Since $\eta$ is even under $ Z_{2}^{{\rm acc}}$, the other daughter must be $ Z_{2}^{{\rm acc}}$-odd,
\begin{equation}
\eta \to X'\phi_{R},~~~\eta \to X'\phi_{I},~~~\eta \to X'^{*}X'^{*}~. \label{eq:etaDecay}
\end{equation}
The $ Z_{2}^{{\rm acc}}$ symmetry excludes $X'\zeta$, $X'h$, and $X'X_{3}$ as decay modes.

\item{\textit{Decays of $X'$.}} Due to its $Z_{3}^{D}$ charge, an unstable $X'$ must have $\eta$ among its decay products. Also, as a carrier of $ Z_{2}^{{\rm acc}}$, $X'$ must also decay to $ Z_{2}^{{\rm acc}}$-odd state,
\begin{equation}
X'\to \eta \phi_{R},~~~X'\to \eta \phi_{I}~. \label{eq:XpDecay}
\end{equation}
The $\zeta$, $h$, and $X_{3}$ cannot be final products since these are all even under $ Z_{2}^{{\rm acc}}$.

\item{\textit{Decays of $\phi_{R}$.}} Since this scalar only carries $ Z_{2}^{{\rm acc}}$ charge, if unstable it must decay to $\phi_{I}$ or $X'$. When $\phi_{I}$ is a daughter, the other state must be neutral under both Abelian groups,
\begin{equation}
\phi_{R}\to \phi_{I}X_{3},~~~\phi_{R}\to X'\eta^{*}~. \label{eq:Rdecay}
\end{equation}

\item{\textit{Decays of $\phi_{I}$.}} The decays of $\phi_{I}$ mirrors those of $\phi_{R}$ because both carry the same charges,
\begin{equation}
\phi_{I}\to \phi_{R}X_{3},~~~\phi_{I}\to X'\eta^{*}~. \label{eq:Idecay}
\end{equation}

\item{\textit{Decays of $\zeta$.}} By mixing with the $h$, the $\zeta$ ends up with SM-like Higgs decay modes, and it is always unstable as long as $\lambda_{H\Phi}\neq0$.

\item{\textit{Decays of $X_{3}$.}} As a singlet under $Z_{3}^{D}\times  Z_{2}^{{\rm acc}}$, the only possible decay modes of $X_{3}$ are
\begin{equation}
X_{3}\to \phi_{R}\phi_{I},~~~X_{3}\to \eta\eta^{*}. \label{eq:X3decay}
\end{equation}
The decay $X_{3}\to X'X'^{*}$, while allowed by the quantum numbers, is kinematically forbidden because the mass relation $M_{3}=\sqrt{3}M_{1}$ guarantees $M_{3}<2M_{1}$. In addition, as a gauge boson whose longitudinal mode is a linear combination of quadruplet pseudoscalars (see \ref{sec:appB}), $X_{3}$ is forbidden to decay into the CP-even final states $\phi_{R}\phi_{R}$, $\phi_{I}\phi_{I}$, $hh$, $h\zeta$, or $\zeta \zeta$.
\end{itemize}
It can be concluded that all bosons with nontrivial $Z_{3}^{D}\times  Z_{2}^{{\rm acc}}$ quantum numbers decay into each other. 

\subsection{Kinematic (in)stability conditions\label{subsec:unstab}}
The kinematic instability conditions on the whole dark sector can be inferred from the decays listed in the previous section. Notice that only the minimal kinematic condition needs to be satisfied to render the parent particle unstable,
\begin{align}
\phi_{R}:~~~M_{R} &> \text{Min}\{ M_{I}+M_{3},~M_{\eta}+M_{1} \} \\
\phi_{I}:~~~M_{I} &> \text{Min}\{ M_{R}+M_{3}~,M_{\eta}+M_{1} \} \\
\eta:~~~M_{\eta}  &> \text{Min}\{ 2M_{1},~M_{R}+M_{1},~M_{I}+M_{1} \} \\
   X':~~~M_{1} &> \text{Min}\{ M_{R}+M_{\eta},~M_{I}+M_{\eta} \} \\
   X_{3}:~~~M_{3} &> \text{Min}\{ M_{R}+M_{I},~2M_{\eta} \}~. \label{eq:nameordering}
\end{align}
In order to avoid establishing a dark sector with too many stable states (whether from quantum numbers or from kinematic means), we will always seek the instability of $X_{3}$. The minimal decay condition of $X_{3}$ is dependent on the $\phi_{R}$, $\phi_{I}$ and $\eta$ mass ordering. In the notation of Eq. (\ref{eq:nameordering}),
\begin{align}
M_{3} &\stackrel{I\eta R}{>} \begin{cases} 2M_{\eta}, & \text{if }M_{I}<(1+\tfrac{\sqrt{2}}{2}M_{\eta}) \\ M_{R}+M_{I}, & \text{ if }M_{I}>(1+\tfrac{\sqrt{2}}{2}M_{\eta})~. \end{cases}  \\
M_{3} &\stackrel{IR\eta}{>} 2M_{\eta}~, \\
M_{3} &\stackrel{RI\eta}{>} 2M_{\eta}~, \\
M_{3} &\stackrel{R\eta I}{>} \begin{cases} M_{R}+M_{I}, & \text{if }M_{I}<(1-\tfrac{\sqrt{2}}{2}M_{\eta}) \\ 2M_{\eta}, & \text{ if }M_{I}>(1-\tfrac{\sqrt{2}}{2}M_{\eta}) \end{cases}
\end{align}
where $(1-\sqrt{2})/2)M_{\eta}\approx 0.29M_{\eta}$ and $(1+\sqrt{2})/2)M_{\eta}\approx 1.71M_{\eta}$. These conditions are obtained, in each mass ordering, by finding which is the smallest of $M_{R}+M_{I}$ and $2M_{\eta}$ in the allowed range of $M_{I}/M_{\eta}$. The interplay between sets of decay conditions is detailed in Appendix \ref{sec:appC}, and in the next subsection we provide a summary of such interplay.

\subsection{Viable multi-component DM scenarios\label{subsec:viable}}

The different mass orderings for the spectrum discussed earlier lead up different dark matter scenarios. These can be classified by the number of (effectively) stable DM candidates. Given the mass relations in Eqs.~(\ref{eq:gaugeMassRel}) and Eq. (\ref{eq:scalarMassRel}), it is expected that some scenarios cannot be realized consistently due to kinematic considerations, i.e. the compressed nature of the scalar and gauge mass spectra. We looked at the class of scenarios determined by the $\phi_{R}$, $\phi_{I}$, and $\eta$ mass ordering (\ref{eq:nameordering}), and within each, determined valid mass ranges for $M_{3}$ and $M_{1}$ within the hierarchy of scalar masses. One must note that $\zeta$ plays no role in classifying these mass orderings because, as commented in Sec. \ref{subsec:decays}, it does not appear as a daughter particle in any of the tree-level, two body decays listed earlier. Naively, the presence of two discrete symmetries suggests two DM candidates, namely the lightest state charged under each of the discrete groups. However, in this section we show that in order to satisfy the kinematical constraints a consistent scenario actually leads to three candidates. Below we follow the kinematical arguments that lead to this situation. First, the unviability of a two-component setup is explained in virtue of such conditions. Next, we describe how going to three-component relaxes the previous constraint, and finally, argue that for  one-component the constraint is even stronger, making it unviable too.
\\

\textit{Two-component DM.} A priori, the $Z_{3}^{D}\times  Z_{2}^{{\rm acc}}$ symmetry seems to accommodate a multitude of possibilities for two-component DM in virtue of the two different quantum numbers involved. The specific instability conditions for each case are tabulated in Appendix \ref{sec:appC}, where it is shown how they are mutually inconsistent in every ordering of $M_{R}$, $M_{I}$, and $M_{\eta}$. We found that all two-component scalar DM scenarios with $X'$ (and hence $X_{3}$) heavier\footnote{Heavier yet within an order of magnitude from them.} than the two lightest dark scalars are inconsistent. The reason is that the minimal decay conditions often require the unstable, third heaviest scalar (for example, $\phi_{R}$ in the $R\eta I$ regime, or $\phi_{I}$ in $IR\eta$) to be heavier than the upper bound set by Eqs.~(\ref{eq:regimeMaster}) or by the mass relation in Eq. (\ref{eq:scalarMassRel}). Also, the fate of the scenarios with one scalar and one gauge boson DM candidates is similar: when $X'$ is the lightest state and one of $\phi_{R}$, $\phi_{I}$, or $\eta$ is the next-to-lightest, it is impossible to destabilize all other Beyond-the-Standard Model (BSM) states in the absence of large gauge-scalar mass gaps. This is also true when one of $\phi_{R}$, $\phi_{I}$, or $\eta$ is the lightest DM and $X'$ is the next-to-lightest. 
\\

\textit{Three-component DM.} Having proved the absence of consistent two-component DM scenarios, we now relax the minimal decay conditions on one more dark state with the purpose of identifying self-consistent setups. In these scenarios, in addition to the lightest $Z_{3}^{D}$ and $ Z_{2}^{{\rm acc}}$ states, one dark state is rendered stable by having its decays forbidden kinematically rather than by the discrete symmetry assignments. We find that a viable scenario can be obtained when all three DM candidates are scalars. One may ask if a gauge boson could be a DM candidate along with two scalars. However, in the case where the two lightest dark scalars and $X'$ are the DM candidates, one must ensure that $X_{3}$ is unstable. Unfortunately, simultaneously satisfying the $X_{3}$ and third-heaviest scalar instability conditions turns impossible without invoking large mass gaps.
\\

\textit{One-component DM.} The gauge boson $X'$ is the only state with nontrivial charge assignments under both discrete symmetries. Then, in principle, it can be the only stable DM candidate if lighter than all dark scalars. This scenario is in fact a more restricted version of the two-component DM setup where $X'$ is the lightest state, because now the lightest dark scalar must satisfy a decay condition. Under our two-body decay treatment, this scenario is proved unviable, ruling out the one-component $X'$ DM case.
\\

Summarizing, we found that when $X_{3}$ and $X'$ are the only unstable states, i.e. a three-scalar DM scenario, all instability conditions hold at the same time. Furthermore, this is true in each of the four orderings $R\eta I$, $RI\eta$, $IR\eta$, and $I\eta R$. There are two reasons why this happens: first, the absence of contrived unstability conditions on $\phi_{R}$, $\phi_{I}$, $\eta$, and second, due to a slightly increased flexibility in pushing $X_{3}$ and $X'$ to heavier masses by increasing $g_{D}$ alone. In order to ease the upcoming numerical analysis, we define the mass ratio $r=M_{I}/M_{\eta}$, then the scalar masses look like
\begin{equation}
M_{\eta},~~~M_{I}=rM_{\eta},~~~M_{R}=M_{\eta}\sqrt{3-r^{2}}, \label{eq:scalardark}
\end{equation}
where $0<r<\sqrt{3}$. We also cast the instability condition on $X'$ (either $M_{1}>M_{R}+M_{\eta}$ or $M_{1}>M_{I}+M_{\eta}$, depending on the dark scalar masses regime) as the condition $g_{D}>g_{D}^{*}$, where $g_{D}^{*}$ is the smallest gauge coupling value at which $X'$ is unstable. The complete dark sector mass orderings in the four main regimes, along with the $r$ allowed range and the kinematic condition for the decay of $X'$, are collected in Table \ref{tab:summaryregimes}. 

\setlength
\tabcolsep{6pt} 
\begin{table}[h!]
\centering
\begin{tabular}{|c|c|c|c|} 
\hline
\textbf{Regime} & $\boldsymbol{r}\textbf{ range}$ & \textbf{Mass ordering} & $\boldsymbol{X'}$\textbf{ unstable if }$\boldsymbol{g_{D}>g_{D}^{*}}$\\
\hline
$I\eta R$ & $\sqrt{2}<r<\sqrt{3}$ & $M_{3}>M_{1}>M_{I}>M_{\eta}>M_{R}$ & $g_{D}^{*}=\sqrt{\dfrac{2}{3}}\dfrac{M_{\eta}}{v_{D}}(\sqrt{3-r^{2}}+1)$ \\
 & & or  $M_{3}>M_{I}>M_{1}>M_{\eta}>M_{R}$ &  \\
 \hline
$IR\eta$ & $\sqrt{\dfrac{3}{2}}<r<\sqrt{2}$ & $M_{3}>M_{1}>M_{I}>M_{R}>M_{\eta}$ & $g_{D}^{*}=\sqrt{\dfrac{2}{3}}\dfrac{M_{\eta}}{v_{D}}(\sqrt{3-r^{2}}+1)$ \\
\hline
$RI\eta$ & $1<r<\sqrt{3/2}$ & $M_{3}>M_{1}>M_{R}>M_{I}>M_{\eta}$ & $g_{D}^{*}=\sqrt{\dfrac{2}{3}}\dfrac{M_{\eta}}{v_{D}}(r+1)$ \\
\hline
$R\eta I$ & $0<r<1$ & $M_{3}>M_{R}>M_{1}>M_{\eta}>M_{I}$ & $g_{D}^{*}=\sqrt{\dfrac{2}{3}}\dfrac{M_{\eta}}{v_{D}}(r+1)$ \\
 & & or $M_{3}>M_{1}>M_{R}>M_{\eta}>M_{I}$ & \\
\hline
\end{tabular}
\caption{Summary of the full mass orderings, range of $r=M_{I}/M_{\eta}$, and $X'$ instability condition on $g_{D}$ for the three-component scalar DM scenarios.} \label{tab:summaryregimes}
\end{table}
 It must be noted that in these scenarios, due to an enhanced annihilation within the dark sector, the relative contributions to the observed DM abundance by the three DM species can differ greatly so that they resemble a setup with effectively one or two DM components (we will refer to such cases as  one- and two-component-like setups). For this reason, rather than providing an exhaustive analysis of every case, we limit ourselves to reveal and describe regions of the parameter space where three-component DM is realized with the three abundances contributing comparably to the observed relic density.

\section{Numerical study\label{sec:numerical}}
In this section we select one of the dark scalar mass regimes shown in Table \ref{tab:summaryregimes} and perform a numerical analysis on it. Such analysis is possible with the help of \texttt{Mathematica}'s model generator \texttt{SARAH v4.15.1} \cite{Staub:2013tta}, the spectrum calculator \texttt{SPheno v4.0.5} \cite{Porod:2011nf}, and the C++-based dark matter analyzer \texttt{MicrOmegas v6.0.5} \cite{Alguero:2023zol}, which computes the DM abundance and contributions by integrating the coupled system Boltzmann equations of the DM candidates.

The two regimes where the dark scalar masses are the most compressed are the $RI\eta$ and $IR\eta$ regimes, then the value of any of the dark scalar masses acts as a ballpark estimate for the other two masses. Based on this observation, we choose the $RI\eta$ regime for the upcoming analysis. The starting point of for our analysis of three-component DM is a look at the available free parameters, namely
\begin{equation}
g_{D},~v_{D},~r,~x,~~M_{\zeta},~M_{\eta},~\text{and}~\lambda_{H\Phi}~.
\end{equation}
In this set, we choose to fix $g_{D}$, $v_{D}$, $r$, $x$, and $M_{\zeta}$, and scan over ranges of $M_{\eta}$ and the Higgs portal coupling $\lambda_{H\Phi}$. The gauge coupling will be taken above the $g_{D}^{*}$ value shown in the last column of Table \ref{tab:summaryregimes}, which is $M_{\eta}$-dependent. Meanwhile, the dark VEV will be fixed at 1 TeV. As indicated in the previous section, the value of $r=M_{I}/M_{\eta}$ within the $0<r<\sqrt{3}$ range fixes the mass ordering between $\phi_{R}$, $\phi_{I}$ and $\eta$. The $r$ ratio will be set to the value of the midpoint $\bar{r}$ of that $r$-range, i.e. $\bar{r}=(1+\sqrt{3/2})/2$ in the $RI\eta$ regime. For the quadruplet components, a small mixing parameter $x$ is chosen, $x=0.1$. Another of our assumptions is that the SM Higgs is the lightest of two $Z_{3}^{D}\times  Z_{2}^{{\rm acc}}$ singlets, $M_{h}<M_{\zeta}$. This is ensured, for example, with the arbitrary choice $M_{\zeta}^{2}\approx 3\times10^{4}\text{ GeV}^{2}$. The dark scalar masses will be parametrized in terms of $M_{\eta}$ and scanned over the tens of GeV up to a few TeV\footnote{For multi-component DM in the sub-GeV regime, see Ref. \cite{Betancur:2021ect}. }. Meanwhile, $\lambda_{H\Phi}$ will be varied from $10^{-5}$ to $10^{-2}$ to determine regions that barely escape direct detection searches. The scans performed below are possible through \texttt{SARAH}'s extension \texttt{SSP v1.2.5} \cite{Staub:2011dp}.

\begin{figure}
\centering
\includegraphics[scale=0.8]{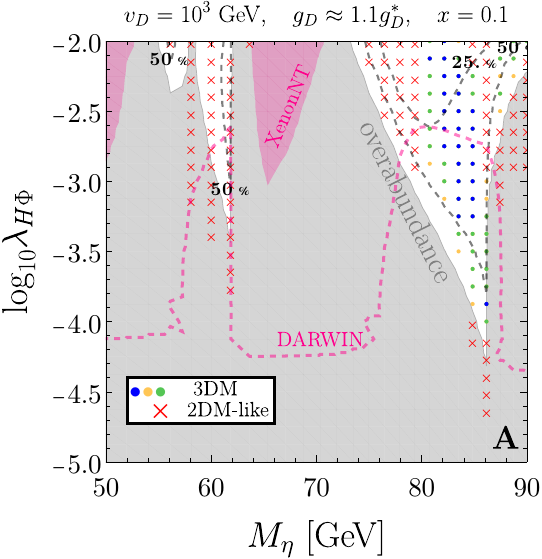}
\hspace{0.5mm}
\includegraphics[scale=0.8]{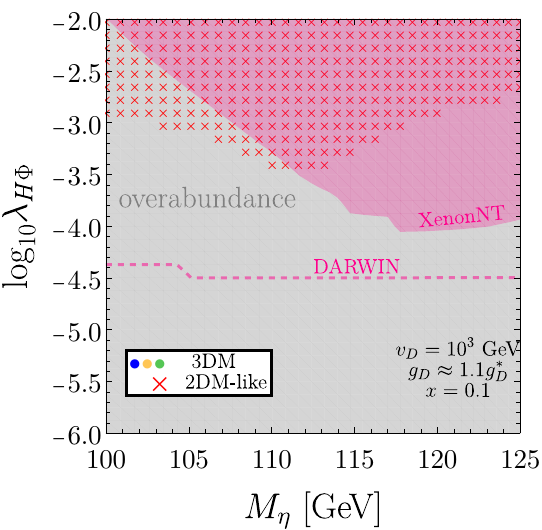}
\hspace{0.5mm}
\includegraphics[scale=0.8]{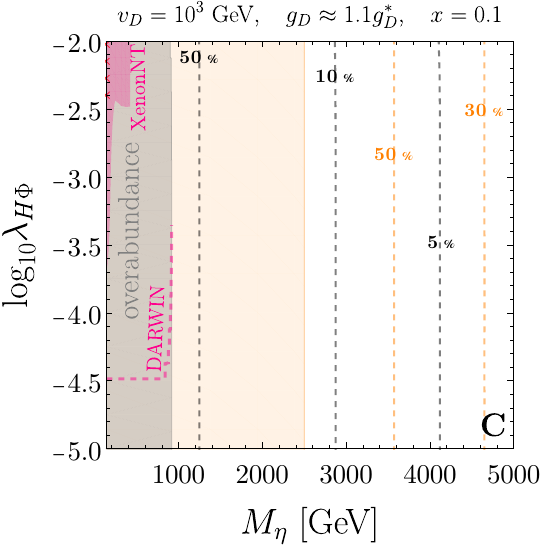}
\caption{
Dark matter abundance and direct detection constraints on the $RI\eta$ regime at fixed dark gauge coupling $g_{D}$, quadruplet VEV $v_{D}$, and quadruplet mixing $x$. The colored markers outline regions with three-component DM (blue, green, and yellow) or two-component-like DM (red crosses). The rest of the regions are one-component-like. Three CP-even scalar mass orderings are shown. \textbf{Panel A}: $M_{R}<M_{h}<M_{\zeta}$, where $g_{D}\sim 0.1$. \textbf{Panel B}: $M_{h}<M_{R}<M_{\zeta}$, with $g_{D}\sim 0.2$. \textbf{Panel C}: $M_{h}<M_{R}<M_{\zeta}$, for which $g_{D}\sim O(1).$
}
\label{fig:RIEscans}
\end{figure}

A representative two-parameter scan in the $RI\eta$ regime is that of Fig. \ref{fig:RIEscans}, whose different panels show different ordering of the $\phi_{R}$ mass (the CP-even dark scalar) with respect to those of $h$ and $\zeta$\footnote{Three panels are shown instead of a single one to avoid numerical artifacts caused by the crossing of CP-even mass eigenvalues in our code. }. The dark gray solid line contains points where the total DM abundance amounts to $100\%$ of the PLANCK measurement, $\Omega h^{2}\approx 0.12$ \cite{Planck:2018vyg}, whereas dashed lines indicate contours of underabundance. Meanwhile, the gray shade covers regions where the total DM exceeds the observed abundance.

In the parameter window shown in panel A of Fig. \ref{fig:RIEscans}, where $M_{R}<M_{h}<M_{\zeta}$, the most prominent features are the two dips where the overabundance exclusion relaxes. Funnels like these are typical of resonant annihilation through $s$-channel Higgs exchange, which explains their occurrence at approximately $M_{h}/2$ and $M_{\zeta}/2$. Only one of the funnels, that of $\zeta$, accommodates a narrow region where $\Omega_{I}\sim\Omega_{\eta}\sim\Omega_{R}$, namely, three-component DM.

In order to graphically indicate the ratios that differentiate between scenarios of distinct effective number of DM species, we first fix the (arbitrary) proportions that they must satisfy to fall in one or another category. Let's call the DM species abundances $\Omega_{1}$, $\Omega_{2}$, and $\Omega_{3}$, where $\Omega_{1}<\Omega_{2}<\Omega_{3}$ and $\Omega_{i}$ is any of $\Omega_{R}$, $\Omega_{I}$, or $\Omega_{\eta}$. One-component-like DM, namely $\Omega_{1}\lesssim \Omega_{2}\ll\Omega_{3}$, satisfies $\Omega_{2}<\tfrac{1}{10}\Omega_{3}$. For this case, the points will not be indicated in any plot, that is, all white regions are meant to be one-component-like DM. For two-component-like DM, namely $\Omega_{1}\ll \Omega_{2}\lesssim \Omega_{3}$ our chosen criteria are $\Omega_{1}<\tfrac{1}{4}\Omega_{2}$ and $\Omega_{2}>\tfrac{1}{10}\Omega_{3}$, which we identify as red crosses in each plot. Finally, three-component DM points are those which satisfy the condition $\Omega_{1} \lesssim \Omega_{2} \lesssim \Omega_{3}$. In this case we consider three different fractions\footnote{These fractions are not mutually exclusive.}: $\Omega_{1}>\tfrac{1}{2}\Omega_{2}$ and $\Omega_{2}>\tfrac{1}{2}\Omega_{3}$ (blue dots), $\Omega_{1}>\tfrac{1}{3}\Omega_{2}$ and $\Omega_{2}>\tfrac{1}{3}\Omega_{3}$ (green dots), and $\Omega_{1}>\tfrac{1}{4}\Omega_{2}$ and $\Omega_{2}>\tfrac{1}{4}\Omega_{3}$ (yellow dots).

Understanding why only one of the funnels in panel A of Fig. \ref{fig:RIEscans} harbors comparable partial abundances for the three scalars requires looking at the main contributions to annihilation there. First, we looked at the $h$ funnel. Having $M_{\eta}$ on the horizontal axis clearly indicates where $\eta$ undergoes resonant $s$-channel annihilation through $h$, dominated by $\eta \eta^{*}\to b\bar{b}$. According to the mass relations in Eq.(\ref{eq:scalardark}), when $\eta$ encounters resonant annihilation near $M_{h}/2$ the heavier $ \phi_{I}$ mass is barely past this value, but still annihilating mostly through $\phi_{I}\phi_{I}\to W^{+}W^{-}$ mediated by $s$-channel Higgs. Likewise, $\eta$ sitting at the Higgs funnel implies that the $\phi_{R}$ mass is past $M_{h}/2$ too. But $\phi_{R}$ ends up a few GeV below $M_{\zeta}/2$, which enhances $\phi_{R}$ annihilation through $s$-channel $\zeta$. Yet, $\phi_{R}$ preferentially undergoes DM conversion into $\eta$ through $\phi_{R}\phi_{R}\to \eta\eta^{*}$, which is possible kinematically as $\phi_{R}$ is heavier than both $\phi_{I}$ and $\eta$. At the $h$-funnel, the dominant cross sections of $\phi_{I}$ and $\eta$ have been checked to be comparable within one order, as opposed to the interconversion cross section of $\phi_{R}$ which is about one hundred times larger. Thus, $\Omega_{R}$ is relatively suppressed with respect to $\Omega_{I}$ and $\Omega_{\eta}$ around $M_{h}/2$, and an effectively two-component DM scenario is rendered there (see red crosses). Next, we turn our attention to the $\zeta$-funnel. When $\eta$ sits at around $M_{\zeta}/2$, its annihilation is resonant, with dominant channel $\eta\eta^{*}\to W^{+}W^{-}$. At that $M_{\eta}$ value, the heavier $\phi_{I}$ is past its own $\zeta$ resonant annihilation but is still dominated by $\phi_{I}\phi_{I}\to W^{+}W^{-}$ with $\zeta$ exchange in the $s$-channel. By the same reason, $\phi_{R}$ is also past its $\zeta$ resonant peak and undergoes conversion to $\eta$ that, while dominated by $\phi_{R}\phi_{R}\to W^{+}W^{-}$, is closely followed by conversion to $\phi_{R}\phi_{R}\to \eta \eta^{*}$ through $t$-channel $X'$ exchange. We have checked that the cross sections for these annihilation channels of $\eta$, $\phi_{I}$ and $\phi_{R}$ are comparable within an order of magnitude when $M_{\eta}$ is around $M_{\zeta}/2$. As a result, $\Omega_{\eta}\sim \Omega_{I}\sim \Omega_{R}$ in that region.

In panel B of Fig. \ref{fig:RIEscans}, the dark scalar $\phi_{R}$ is found between the two Higgses, $M_{h}<M_{R}<M_{\zeta}$. At these $M_{\eta}$ values the $h$ and $\zeta$ funnels have degraded and annihilation is no longer resonant. While $\eta$ dominates the DM abundance, it is closely followed by $\phi_{I}$ (as depicted by the red crosses) provided that the portal coupling is large enough. Annihilation of $\eta$ and $\phi_{I}$ is still determined by Higgs portal interactions, leading to $\eta \eta^{*}\to VV^{*}$ and  $\phi_{I} \phi_{I}\to VV^{*}$ with $V=W,Z$. These\textbf{} are mediated by $h$ and $\zeta$ in the $s$-channel and have comparable size. Yet, the absence of resonant annihilation rules out the entire window by overabundance considerations.

Panel C of Fig. \ref{fig:RIEscans} corresponds to a $\phi_{R}$ heavier than the two Higgses, $M_{h}<M_{\zeta}<M_{R}$. In this region $\eta$ comprises most of the relic, effectively a one-component-like DM scenario, and the PLANCK measurement is fulfilled at a near-TeV $M_{\eta}$. The (non-resonant) annihilation of $\eta$ is driven by $\eta \eta^{*}\to \zeta \zeta$ through a four-point vertex, followed by dark-to-dark conversion $\eta \eta^{*}\to \phi_{I} \phi_{I}$ via $X'$ exchange in the $t$-channel. Since the $\eta$ annihilation stops being determined by the Higgs portal coupling, the relic contours become nearly vertical. 

To continue our analysis, we discuss the role of dark matter semi-annihilation in the scans above. The presence of dark states $\varphi$ carrying a $Z_{3}^{D}$ quantum charge introduces annihilations of the type $\varphi \varphi \to \varphi^{*}\chi$ where $\chi$ is $Z_{3}^{D}$-neutral (dark or visible). In this model $\varphi$ is either $\eta$ or $X'$, but it is enough to consider $\eta$ as it is the lighter of the two. In the region near the $h$ and $\zeta$ funnels, annihilation of the dark scalars ($\eta$ included) is overwhelmingly driven by resonant $s$-channel self-annihilation through $h$ and $\zeta$, as discussed earlier. While these $\eta \eta^{*}\to \text{SM }\text{SM}$ channels decrease in size away from the funnels, they still dominate annihilation as $M_{\eta}$ varies in the hundred-GeV ballpark. This is the case, for instance, in the region of panel B of Fig. \ref{fig:RIEscans}, which is nevertheless excluded due to DM relic overabundance. Further away from the funnels, at even heavier $M_{\eta}$, DM becomes only one-component-like, dominantly $\eta$. This is indeed the case of the region shown in panel C of Fig. \ref{fig:RIEscans}. There, at the selected parameters, semi-annihilation remains subdominant due a ratio of thermal cross sections $\langle v\sigma\rangle_{\text{self}}/\langle v\sigma\rangle_{\text{semi}}\approx 4$ at the decoupling temperature, as checked with the help of \texttt{MicrOmegas}. To depict the predominance of self-annihilation in that region, the same overabundance shade and underabundance contours are repeated in orange but now with $\eta$ self-annihilation turned off~\footnote{This depiction is preferred over showing the contours with semi-annihilation switched off because the effect is barely discernible.}. The effect of switching off self-annihilation is manifested as a considerable shift towards larger $M_{\eta}$ in said contours.

In order to grasp the extent to which $\eta \eta\to \eta^{*}\zeta$ can be rendered comparable to $\eta \eta^{*}\to \zeta \zeta$ in our regions of study, we fixed the $\lambda_{H\Phi}$ quartic and let $M_{\zeta}$ vary while keeping the rest of parameters at the same values as before.
\begin{figure}
\centering
\includegraphics[scale=0.8]{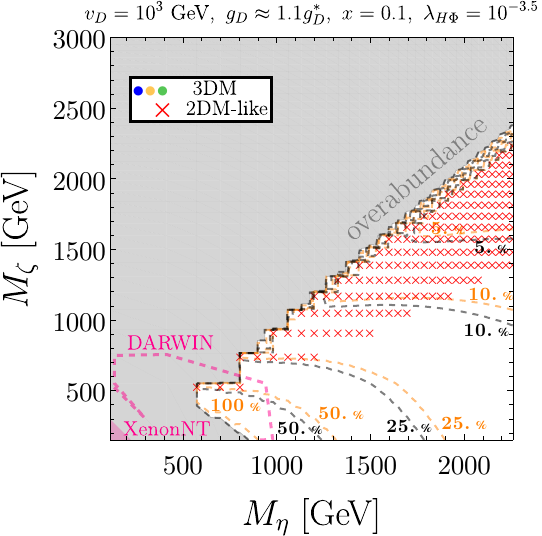}
\caption{The $M_{\eta}$ and $M_{\zeta}$ interplay in the hundred-GeV to few-TeV region, corresponding to panel C of Fig. \ref{fig:RIEscans} but now at fixed Higgs portal quartic. The dashed contours indicate underabundance values with all channels active (gray) and without $\eta \eta\to \eta^{*}\zeta$ semi-annihilation (orange). The regions below pink contours are excluded by direct detection considerations.}
\label{fig:SelfSemiComparison}
\end{figure}
In figure \ref{fig:SelfSemiComparison}, $M_{\zeta}$ is varied from hundreds of GeV to a few TeV, which includes regions where the ratio $M_{\zeta}/M_{\eta}$ approaches 1. Underabundant DM relic dashed contours are shown in the case when all annihilation channels open (gray) and with semi-annihilation turned off (orange). We have checked that $\eta$ semi-annihilation only becomes comparable to self-annihilation at $M_{\zeta}/M_{\eta}>1$ ratios, which for this window lies within the overabundant exclusion. As for the underabundant region, the effect of semi-annihilation is subdominant. Regardless, here, three regions are discerned: in the region where $\zeta$ is much lighter than $\eta$, at fixed $M_{\eta}$ we notice a soft variation of the underabundance contours as $M_{\zeta}$ is varied. In fact, the contours become more vertical towards the smallest $M_{\zeta}$ value shown, and start being insensitive to it. A second region is that where contours are almost horizontal, where $M_{\zeta}$ becomes closer to $M_{\eta}$. Despite being marked as a two-component-like region, we have checked that $\eta$ is the most abundandant of the two species. Here, although $\eta$ is already past its self-annihilation resonant peak at $M_{\zeta}/2$, higher $M_{\zeta}$ brings this peak slightly closer to $M_{\eta}$, causing  an increase of self-annihilation and a decrease in the $\eta$ relic. A third region is that where $M_{\zeta}/M_{\eta}$ is very close to 1. At fixed $M_{\eta}$, heavier $M_{\zeta}$ starts limiting the phase space available for $\eta\eta^{*}\to \zeta \zeta$, which lowers this cross section and leaves higher DM abundance. 

\begin{figure}[h!]
\centering
\includegraphics[scale=0.8]{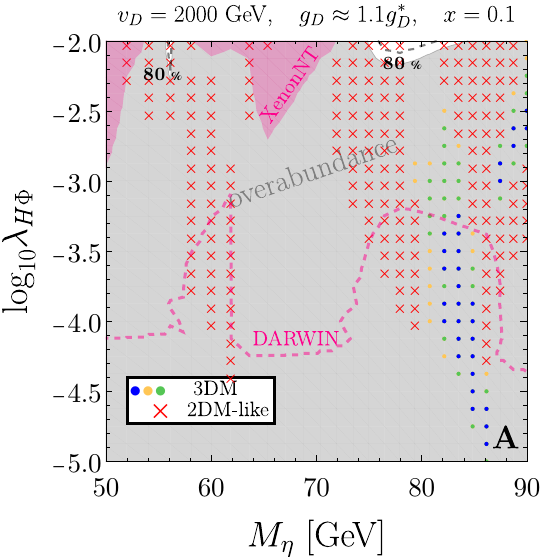}
\hspace{0.5mm}
\includegraphics[scale=0.8]{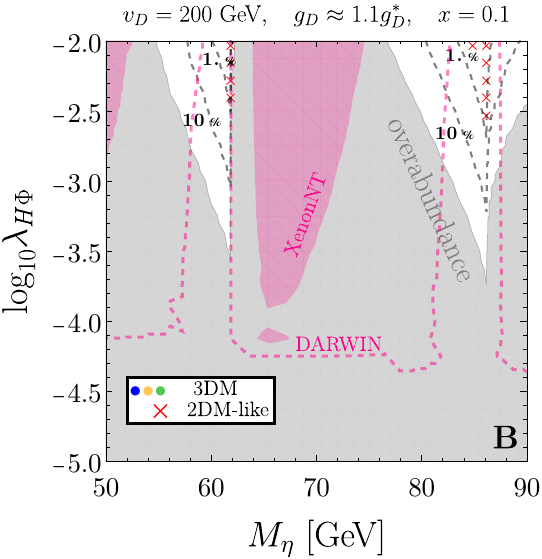}
\hspace{0.5mm}
\includegraphics[scale=0.8]{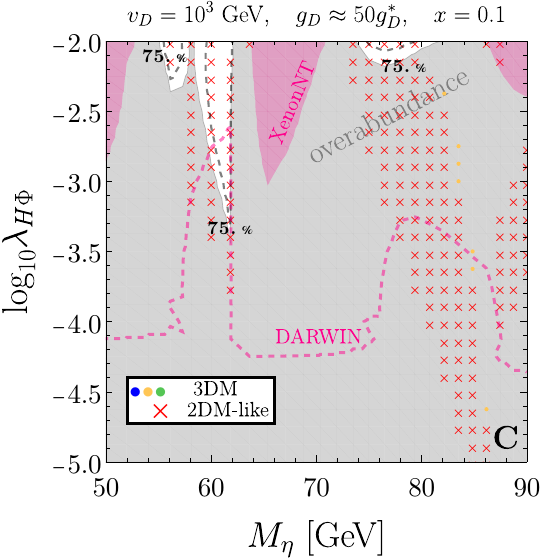}
\caption{Scans analogue to panel A of Fig. \ref{fig:RIEscans}, now with different $v_{D}$ (panels A and B) and $g_{D}$ (panel C). In panels A and B, $g_{D}\sim 0.5$, while $g_{D}\sim O(1)$ in panel C.}
\label{fig:otherCuts}
\end{figure}

In the spirit of complementarity, the sensitivity of the three-component DM regions to the choices of benchmark parameters in the windows above is shown for different choices of the main dark parameters, namely the dark gauge coupling and dark VEV, in the panels of Fig. \ref{fig:otherCuts}.
In panel A of Fig. \ref{fig:otherCuts} one sees that when $v_{D}$ is larger than that in panel A of Fig. \ref{fig:RIEscans}, the $h$ and $\zeta$ funnels erode. This is so because as $v_{D}$ grows while keeping the scalar masses at the same ballpark as before, the interconversion amplitudes between the $\phi_{R}$, $\eta$, and $\phi_{I}$ decrease in size due to $X'$ and $X_{3}$ becoming heavier in the propagators. This causes the region of comparable abundances to be buried by the overabundance shade. The opposite effect, i.e. deeper funnels, is achieved with a $v_{D}$ smaller than the one chosen earlier, as shown in panel B of Fig. \ref{fig:otherCuts}. The tradeoff is, however, the preponderance of one-component-like points with acceptable DM abundance. Finally, for a $g_{D}$ choice much larger than the one used before, the funnels deteriorate once again as shown in panel C of Fig. \ref{fig:otherCuts}. This is due to enhanced DM interconversion via $X'$ and $X_{3}$ exchange, which relies on gauge-scalar vertices.

To conclude this section, a description of dark matter signals from nuclear-recoil based scattering, i.e. direct detection, is presented. The panels of Fig. \ref{fig:RIEscans} illustrate the comparison between the spin-independent (SI) cross section of the model at the chosen parameters against the XenonNT \cite{XENON:2023cxc} limit (magenta shade) and the future sensitivity of the DARWIN collaboration \cite{DARWIN:2016hyl} (magenta dashed line). Due to the presence of three DM species, the constraint imposed by these experiments on the $i$-th species is not on its SI cross section $\sigma_{i}^{\text{SI}}$ but on its SI cross section rescaled by its relative abundance with respect to the observed value\footnote{This is so because experimental limits assume the PLANCK abundance value in the recoil rate calculation.}, $\sigma_{i}^{\text{SI}}\times \Omega_{i}/\Omega_{\text{PLANCK}}$. For panel A of Fig. \ref{fig:RIEscans}, which is the only one with a region where all three abundances are comparable, we opted to compare the XenonNT and DARWIN limits with the weighted sum
\begin{equation}
\sum_{i=R,I,\eta}\dfrac{\Omega_{i}}{\Omega_{\text{PLANCK}}}\sigma_{i}^{\text{SI}}~. \label{eq:compositeSI}
\end{equation}
Away from the $\zeta$ funnel the three abundances start differing and the comparison with the sum above can still be done, if taken as a conservative bound. As observed in that figure, the DARWIN projected sensitivity rules out parts of the Higgs and $\zeta$ funnels, but still leaves points that are 3DM while hitting the observed DM abundance. For panels B and C in Fig. \ref{fig:RIEscans}, the direct detection limit is applied to $\sigma_{\eta}^{\text{SI}}\times \Omega_{\eta}/\Omega_{\text{PLANCK}}$ directly because $\eta$ is by far the dominant species there.

\section{Conclusions\label{sec:conclusions}}
An analysis has been presented of a BSM model in order to determine its viability as a multi-component dark matter scenario. The model involves an additional (to the SM gauge symmetry) $SU(2)_D$ local symmetry spontaneously broken to a $Z_3^D$ discrete subgroup through the non-zero VEV a quadruplet scalar. Furthermore, there is an accidental $Z_2^{\rm acc}$ in the broken phase of the model such that, effectively, one has a $SU(2)_D \to Z_3^D \times Z_3^{\rm acc}$ scenario. 

The mass spectrum of the new $SU(2)_D$, i.e. gauge bosons and scalar components of the quadruplet, render interesting kinematic constraints which are found to restrict the possibilities for a viable set of DM candidates. We find that together with the presence of the two discrete symmetries involved, the kinematic constraints only allow for a viable three-component dark matter scenario. 

A thorough numerical study has been performed that shows the regions of parameter space where the model provides a DM abundance consistent with current measurements. The individual contributions from the three DM candidates to the total abundance is shown to vary throughout the parameter in a way that, effectively, one can have regions of parameter space where all three candidates contribute significantly, hence called three-component DM scenario, as well as regions in which only one or two of the three contribute do so, hence called one- and two-component-like scenarios. 

We also conclude that next generation direct detection searches, depending on specific parameter choices that take advantage of resonant DM annihilation, have the potential to find members of a multi-component dark sector. This represents a departure from the familiar expectation to find a unique DM particle candidate.

\section*{Acknowledgments\label{sec:ack}}
Work supported by ANID-Chile under the grant FONDECYT Regular No. 1241855.
C. A. and A. A. thank the Universidad Cat\'olica del Norte and the Universidad de Antofagasta for their hospitality during part of the completion of this work. C.B. acknowledges the Universidad de Colima hospitality where part of this work was carried out.

\appendix
\section{Appendix: The quadruplet potential\label{sec:appA}}
A $J=3/2$ quadruplet  $\Phi$ can be represented either in terms of the $J_{z}$ eigenvalues or as a fully symmetric 3-index tensor with components $\phi^{ijk}$. The correspondence between entries in each representation follows\textbf{}
\begin{equation}
\Phi=\left(\begin{array}{c}
\phi_{3} \\
\phi_{2} \\
\phi_{1} \\
\phi_{0} \\
\end{array}\right)=
\left(\begin{array}{c}
\phi^{111} \\
\sqrt{3}\phi^{112} \\
\sqrt{3}\phi^{122} \\
\phi^{222} \\
\end{array}\right).
\label{eq:quadrupletPhi}
\end{equation}
As shown in Ref. \cite{Borah:2022dbw}, the scalar potential of $\Phi$ appearing in Eq.~(\ref{eq:potentialHPhi}) has the structure\textbf{}
\begin{align}
V(&\Phi)
=-\mu_{\Phi}^{2}\sum_{k}\phi_{k}^{\dag}\phi_{k}+\dfrac{\lambda_{1}}{2}\biggl( \sum_{k}\phi_{k}^{\dag}\phi_{k} \biggr)^{2} \notag \\
&- \dfrac{10\lambda_{2}}{9}\biggl[ (\Phi\widetilde{\Phi})_{\boldsymbol{3}}(\Phi\widetilde{\Phi})_{\boldsymbol{3}} \biggr]_{\boldsymbol{1}} - \dfrac{5\lambda_{3}}{18}\biggl[ (\Phi\Phi)_{\boldsymbol{3}}(\Phi\Phi)_{\boldsymbol{3}}+\text{h.c.}\biggr]_{\boldsymbol{1}} - \dfrac{5\lambda_{4}}{9}\biggl[ (\Phi\Phi)_{\boldsymbol{3}}(\Phi\widetilde{\Phi})_{\boldsymbol{3}}+\text{h.c.}\biggr]_{\boldsymbol{1}}
\end{align}
where the components of $\widetilde{\Phi}$ are defined through $\widetilde{\Phi}^{ijk}=\varepsilon^{ii'}\varepsilon^{jj'}\varepsilon^{kk'}\Phi_{i'j'k'}^{*}$ (indices refer to the tensorial notation). The objects $(\Phi \Phi)_{\boldsymbol{3}}$ and $(\Phi \widetilde{\Phi})_{\boldsymbol{3}}$ are $J=1$ objects whose $J_{z}=+1,0,-1$ components are determined by the Clebsch-Gordan coefficients in a $\boldsymbol{3/2}\otimes\boldsymbol{3/2}$ product, and so on. Other contractions of $\Phi$ with higher or lower $J$ are either vanishing, or equivalent to the $J=1$ products included above.


\section{Appendix: Gauge and scalar mixing\label{sec:appB}}

We list in this appendix the algebra of gauge and scalar mixing. In the gauge sector, mass mixing with the $Z$ is absent due to the lack of scalars simultaneously charged under both dark and SM symmetries. Likewise, kinetic mixing is absent from the non-Abelian nature of $SU(2)_{D}$. The mass matrices of the scalars display the full mixing after the spontaneous breaking of the $SU(2)_{D}$ and electroweak symmetries.

The dark scalar-gauge interaction is encoded in the gauge covariant derivative of the quadruplet,
\begin{equation}
D_{\mu}\Phi=\partial_{\mu}\Phi-ig_{D}X_{\mu}^{a}T^{a}\Phi,
\end{equation}
where $a=1,2,3$ (sum of repeated indices is implicit). The $T^{a}$ are $SU(2)_{D}$ generators in the (four-dimensional) $J=3/2$ representation,
\begin{equation*}
J_{1}=\dfrac{1}{2}\left(\begin{array}{cccc}
         0 &  \sqrt{3/2}  & 0 & 0 \\
\sqrt{3/2} & 0 & \sqrt{3/2} & 0 \\
         0 & \sqrt{3/2} & 0 & \sqrt{3/2} \\
         0 & 0 & \sqrt{3/2} & 0
\end{array}\right),~~~~~
J_{2}=\dfrac{1}{2}\left(\begin{array}{cccc}
         0 &  \sqrt{3/2}  & 0 & 0 \\
-\sqrt{3/2} & 0 & \sqrt{3/2} & 0 \\
         0 & -\sqrt{3/2} & 0 & \sqrt{3/2} \\
         0 & 0 & -\sqrt{3/2} & 0
\end{array}\right)
\end{equation*}
\begin{equation*}
J_{3}=\left(\begin{array}{cccc}
3/2 &   0 & 0    & 0 \\
  0 & 1/2 & 0    & 0 \\
  0 &   0 & -1/2 & 0 \\
  0 &   0 &    0 & -3/2
\end{array}\right).
\end{equation*}
The $SU(2)_{D}$ breaking pattern $\langle \Phi\rangle=(1/\sqrt{2})(v_{3},0,0,v_{0})$ induces the mixing matrix for the dark gauge bosons
\begin{equation}
\mathcal{M}_{X}^{2}=
\left(\begin{array}{ccc}
\dfrac{3}{2}g_{D}^{2} & 0 & 0 \\
                    0 & \dfrac{3}{2}g_{D}^{2} & 0 \\
                    0 & 0 & \dfrac{9}{2}g_{D}^{2}
\end{array}\right)v_{D}^{2}, \label{eq:mixVectors}
\end{equation}
in the $\{ X_{1},X_{2},X_{3}\}$ basis. The $X_{1}$ and $X_{2}$ turn out degenerate and can be combined into a complex vector $X'=(X_{1}- iX_{2})/\sqrt{2}$ and its charge-conjugate $X'^{*}=(X_{1}+ iX_{2})/\sqrt{2}$.
\\

In the basis $\{ \sigma_{h},\sigma_{0},\sigma_{3}\}$ of imaginary parts of $h$, $\phi_{0}$ and $\phi_{3}$, the pseudoscalar mixing matrix post-minimization reads
\begin{align}
&\mathcal{M}_{\text{CP-odd}}^{2}= \notag \\
& \left(\begin{array}{ccc}
0 & 0 & 0 \\
0 & \tfrac{1}{2}(-2\lambda_{2}-3\lambda_{3}+(-2\lambda_{2}+\lambda_{3})c_{4\theta}^{-1})s_{\theta}^{2} &  \tfrac{1}{4}(-2\lambda_{2}-3\lambda_{3}+(-2\lambda_{2}+\lambda_{3})c_{4\theta}^{-1})s_{2\theta} \\
0 & \tfrac{1}{4}(-2\lambda_{2}-3\lambda_{3}+(-2\lambda_{2}+\lambda_{3})c_{4\theta}^{-1})s_{2\theta}    & \tfrac{1}{2}(-2\lambda_{2}-3\lambda_{3}+(-2\lambda_{2}+\lambda_{3})c_{4\theta}^{-1})c_{\theta}^{2}
\end{array}\right)v_{D}^{2}. \label{eq:mixCPodd}
\end{align}
where $c_{\theta}$ and $s_{\theta}$ are sine and cosine of $\theta$, respectively, while $c_{4\theta}^{-1}$ stands for the secant of $4\theta$.
\\

Next, the (symmetric) mixing matrix of CP-even states in the basis $\{ \rho_{h},\rho_{0},\rho_{3}\}$ of real parts of $h$, $\phi_{0}$ and $\phi_{3}$ displays the following entries after minimization
\begin{align}
(\mathcal{M}_{\text{CP-even}}^{2})_{1,1} &= 2\lambda_{h}v^{2}, \\
(\mathcal{M}_{\text{CP-even}}^{2})_{2,2} &= \tfrac{v_{D}^{2}}{4}\bigl( 
2(4\lambda_{1}+2\lambda_{2}+\lambda_{3})c_{\theta}^{2}+(2\lambda_{2}-\lambda_{3})(-1+3c_{2\theta})c_{4\theta}^{-1}  \bigr), \\
(\mathcal{M}_{\text{CP-even}}^{2})_{3,3} &= \tfrac{v_{D}^{2}}{4}\bigl( 
4\lambda_{1}+2\lambda_{2}+\lambda_{3}+3(2\lambda_{2}-\lambda_{3})c_{4\theta}^{-1}  \bigr)s_{2\theta}, \\
(\mathcal{M}_{\text{CP-even}}^{2})_{1,2} &= 2\lambda_{H\Phi}vv_{D}c_{\theta}, \\
(\mathcal{M}_{\text{CP-even}}^{2})_{1,3} &= 2\lambda_{H\Phi}vv_{D}s_{\theta}, \\
(\mathcal{M}_{\text{CP-even}}^{2})_{2,3} &= \tfrac{v_{D}^{2}}{4}\bigl( 
-2(4\lambda_{1}+2\lambda_{2}+\lambda_{3})s_{\theta}^{2}-(2\lambda_{2}-\lambda_{3})(1+3c_{2\theta})c_{4\theta}^{-1}  \bigr).
\label{eq:mixCPeven}
\end{align}

Finally, the $2\times2$ complex scalar mixing matrix written in the basis $\{ \phi_{1},\phi_{2} \}$ reads
\begin{equation}
\mathcal{M}_{\text{complex}}^{2}=-\dfrac{1}{12}v_{D}^{2}
\bigl( 2\lambda_{2}+3\lambda_{3}+3(2\lambda_{2}-\lambda_{3}) \bigr)c_{4\theta}^{-1}
\left(\begin{array}{cc}
2s_{\theta}^{2} & s_{2\theta} \\
s_{2\theta}     & 2c_{\theta}^{2}
\end{array}\right)v_{D}^{2}. \label{eq:mixComplex}
\end{equation}
The massless modes of the CP-odd and complex scalar mixing matrices in Eqs. (\ref{eq:mixCPodd}) and (\ref{eq:mixComplex}) that become the longitudinal components of $Z$, $X_{3}$, and $X'$ are, respectively, $G_{Z}=\sigma_{h}$, $G_{X_{3}}=-c_{\theta}\sigma_{0}+s_{\theta}\sigma_{3}$ and $G_{X'}=-c_{\theta}\phi_{1}+s_{\theta}\phi_{2}$.


\section{Appendix: New states instability conditions\label{sec:appC}}

\noindent
\begin{itemize}

\item{$\boldsymbol{I\eta R}$\textbf{ ordering}.}
For this regime, $\sqrt{2}M_{\eta}<M_{I}<\sqrt{3}M_{\eta}$. The decay conditions for $X'$ and $X_{3}$ are 
\begin{align*}
M_{1} &> \text{Min}\{ M_{R}+M_{\eta},~M_{I}+M_{\eta} \}, \\
M_{3} &\stackrel{I\eta R}{>} \begin{cases} 2M_{\eta} & \text{if }M_{I}<(1+\tfrac{\sqrt{2}}{2})M_{\eta} \\ M_{R}+M_{I} & \text{ if }M_{I}>(1+\tfrac{\sqrt{2}}{2})M_{\eta}. \end{cases} 
\end{align*}
For $X'$ the minimal condition is clearly
\begin{equation}
M_{1}>M_{R}+M_{\eta}. \label{eq:IERkincond}
\end{equation}
The two decay modes of $X_{3}$ provide weaker conditions than this one.
The minimal condition for $X'$ is enough to guarantee the instability of both $X'$ and $X_{3}$. Notice that, no matter what, $X_{3}$ is heavier than $\phi_{I}$ because either $M_{3}>2M_{\eta}>M_{I}^{\text{max}}=\sqrt{3}M_{\eta}$ or $M_{3}>M_{R}+M_{I}>M_{I}$. The condition (\ref{eq:IERkincond}) is not strong enough to place $X'$ heavier than $\phi_{I}$, then $M_{1}>M_{I}$ and $M_{I}>M_{1}$ are acceptable. Since (\ref{eq:IERkincond}) implies that $X'$ is separately heavier than $\phi_{R}$ and $\eta$,
\begin{equation*}
M_{3}>M_{I}>M_{1}>M_{\eta}>M_{R}~~~~~\text{or}~~~~~M_{3}>M_{1}>M_{I}>M_{\eta}>M_{R},
\end{equation*}
leaving $\phi_{I}$ as either the third or fourth heaviest odd $ Z_{2}^{{\rm acc}}$ particle. More specifically, the reason two orderings exist is because $M_{R}+M_{\eta}$ can take values in $[M_{\eta},2M_{\eta}]$ but $M_{I}$ lives in $[\sqrt{2}M_{\eta},\sqrt{3}M_{\eta}]$, which is fully contained in the $[M_{\eta},2M_{\eta}]$ range. This regime is self-consistent, with $\phi_{I}$ rendered kinematically stable.

\item{$\boldsymbol{IR\eta}$\textbf{ ordering}.} Defined by $\sqrt{3/2}M_{\eta}<M_{I}<\sqrt{2}M_{\eta}$. The decay conditions for $X'$ and $X_{3}$ are 
\begin{align*}
M_{1} &> \text{Min}\{ M_{R}+M_{\eta},~M_{I}+M_{\eta} \}, \\
M_{3} &\stackrel{IR\eta}{>} 2M_{\eta}.
\end{align*}
For $X'$, the $IR\eta$ regime tells us that $M_{I}+M_{\eta}>M_{R}+M_{\eta}$, then the minimal condition is  
\begin{equation}
M_{1}>M_{R}+M_{\eta} \label{eq:IREkincond}
\end{equation}
For $X_{3}$,
\begin{equation*}
M_{3}=\sqrt{3}M_{1}>M_{\eta}+M_{\eta}~~~\Rightarrow~~~M_{1}>\tfrac{1}{\sqrt{3}}(M_{\eta}+M_{\eta})
\end{equation*}
which is weaker than what (\ref{eq:IREkincond}) implies
\begin{equation*}
M_{1}>M_{R}+M_{\eta}>M_{\eta}+M_{\eta}.
\end{equation*}
It is sufficient to adopt the minimal condition for $X'$. Is $X'$ heavier than the heaviest scalar, $\phi_{I}$? The answer is yes: in the $IR\eta$ regime $M_{I}^{\text{max}}=\sqrt{2}M_{\eta}$ and the inequality right above tells us $M_{1}>2M_{\eta}>M_{I}^{\text{max}}$, then
\begin{equation*}
M_{3}>M_{1}>M_{I}>M_{R}>M_{\eta}~,
\end{equation*}
meaning $\phi_{I}$ is the third heaviest $ Z_{2}^{{\rm acc}}$ particle. This regime is self-consistent, with $\phi_{I}$ rendered kinematically stable by phase space.

\item{$\boldsymbol{RI\eta}$\textbf{ ordering}.} Defined by $M_{\eta}<M_{I}<\sqrt{3/2}M_{\eta}$. The decay conditions for $X'$ and $X_{3}$ are 
\begin{align*}
M_{1} &> \text{Min}\{ M_{R}+M_{\eta},~M_{I}+M_{\eta} \}, \\
M_{3} &\stackrel{RI\eta}{>} 2M_{\eta}.
\end{align*}
For $X'$, in the $RI\eta$ regime $M_{R}+M_{\eta}>M_{I}+M_{\eta}$, then
\begin{equation}
M_{1}>M_{I}+M_{\eta} \label{eq:RIEkincond}
\end{equation}
As for $X_{3}$, 
\begin{equation*}
M_{3}=\sqrt{3}M_{1}>M_{\eta}+M_{\eta}~~~\Rightarrow~~~M_{1}>\tfrac{1}{\sqrt{3}}(M_{\eta}+M_{\eta})
\end{equation*}
but this is weaker than the boxed condition above, which implies
\begin{equation*}
M_{1}>M_{I}+M_{\eta}>M_{\eta}+M_{\eta}.
\end{equation*}
Therefore, it is enough to adopt the minimal condition for $X'$ to simultaneously guarantee the $X'$ and $X_{3}$ decays. Still, we can fix the entire mass ordering: $M_{1}>M_{I}+M_{\eta}>2M_{\eta}>M_{R}^{\text{max}}$ in this regime (which has a value $M_{R}^{\text{max}}=\sqrt{2}M_{\eta}$), then the mass ordering is fixed
\begin{equation*}
M_{3}>M_{1}>M_{R}>M_{I}>M_{\eta}~,
\end{equation*}
We conclude that this regime is self-consistent, with $\phi_{R}$ rendered kinematically stable by phase space.

\item{$\boldsymbol{R\eta I}$\textbf{ ordering}.} Defined by $0<M_{I}<M_{\eta}$. Only $X'$ and $X_{3}$ are required to be unstable. The decay conditions for $X'$ and $X_{3}$ are
\begin{align*}
M_{1} &> \text{Min}\{ M_{R}+M_{\eta},~M_{I}+M_{\eta} \} \\
M_{3} &\stackrel{R\eta I}{>} \begin{cases} M_{R}+M_{I} & \text{if }M_{I}<(1-\tfrac{\sqrt{2}}{2}M_{\eta}) \\ 2M_{\eta} & \text{ if }M_{I}>(1-\tfrac{\sqrt{2}}{2}M_{\eta}) \end{cases}
\end{align*}
For $X'$, in the $R\eta I$ regime clearly $M_{R}+M_{\eta}>M_{I}+M_{\eta}$, then
\begin{equation}
M_{1}>M_{I}+M_{\eta}. \label{eq:REIkincond}
\end{equation}
For $X_{3}$, its two decays give weaker conditions than (\ref{eq:REIkincond}). Notice that no matter what, $X_{3}$ is heavier than $\phi_{R}$,
\begin{equation*}
M_{3}>2M_{\eta}>M_{R}^{\text{max}}~~~~~\text{or}~~~~~M_{3}>M_{R}+M_{I}>M_{R},
\end{equation*}
where $M_{R}^{\text{max}}>\sqrt{3}M_{\eta}\approx 1.73M_{\eta}$. Yet, $X'$ is not necessarily heavier than $\phi_{R}$. Now, the $X'$ decay condition simultaneously implies $M_{1}>M_{I}$ and $M_{1}>M_{\eta}$, so together with the regime ordering, two orderings arise
\begin{equation*}
M_{3}>M_{R}>M_{1}>M_{\eta}>M_{I}~~~~~\text{or}~~~~~M_{3}>M_{1}>M_{R}>M_{\eta}>M_{I}~,
\end{equation*}
which tells that $\phi_{R}$, being stable by phase-space, is not necessarily the third heaviest particle with odd $Z_{2}^{{\rm acc}}$ parity, it can also be the fourth heaviest\footnote{$X'$ is not DM, even when $M_{R}>M_{1}$, because it is unstable.}. We conclude that this regime is self-consistent, with $\phi_{R}$ rendered kinematically stable by phase space.

\end{itemize}

\bibliography{bibliography_Z3Z2}

\end{document}